\documentclass[twocolumn,preprintnumbers,showpacs,aps,prb,superscriptaddress]{revtex4}

\usepackage{graphicx,times,epsfig,color}
\usepackage{float}
\usepackage{amsmath} 
\usepackage{amsthm} 
\usepackage{xfrac}
\usepackage{braket}
\usepackage{amssymb}	
\usepackage{graphics} 
\usepackage{multirow}

\usepackage{comment}

\begin{document}

\def\salto{\vskip 1cm}
\def\lgr{\langle\langle}
\def\rgr{\rangle\rangle}

\title{Non-perturbative effects and indirect exchange interaction between quantum impurities on metallic (111) surfaces}
%
\author{A. Allerdt}
\affiliation{Department of Physics, Northeastern University, Boston, Massachusetts 02115, USA}
\author{R. \v{Z}itko}
\affiliation{Jozef \v{S}tefan Institute, Jamova 39, SI-1000 Ljubljana, Slovenia}
\author{A. E. Feiguin}
\affiliation{Department of Physics, Northeastern University, Boston, Massachusetts 02115, USA}

\begin{abstract}
The (111) surface of noble metals is usually treated as an isolated two dimensional (2D) triangular lattice completely decoupled from the bulk. However, unlike topological insulators, other bulk bands cross the Fermi level. We here introduce an effective tight-binding model that accurately reproduces results from first principles calculations, accounting for both surface and bulk states. We numerically solve the many-body problem of two quantum impurities sitting on the surface by means of the density matrix renormalization group. By performing simulations in a star geometry, we are able to study the non-perturbative problem in the thermodynamic limit with machine precision accuracy. We find that 
there is a non-trivial competition between Kondo and RKKY physics and as a consequence, ferromagnetism is never developed, except at short distances. The bulk introduces a variation in the period of the RKKY interactions, and therefore the problem departs considerably from the simpler 2D case. In addition, screening, and the magnitude of the effective indirect exchange is enhanced by the contributions from the bulk states.
\end{abstract}


\maketitle

\section{Introduction}

Nano-structures assembled on crystal surfaces through single-atom
manipulation with a scanning tunneling microscope (STM) can serve as
model systems for studying magnetism with atomic real-space resolution
as well as excellent energy resolution.
\cite{heinrich2004,otte2009,Spinelli:2015kh} This technique can, 
for example, uncover spatial profiles of magnetic excitations in
artificial one-dimensional spin chains
\cite{hirjibehedin2006,Khajetoorians:2012ji,Spinelli:2014db,Toskovic:2016fv}.
The surface not only supports the spin centres, but also plays a
crucial role in stabilizing magnetic order
\cite{Stepanyuk.Magnetic_nano,Zhou.Strength_and_directionality,
Ignatiev.Magnetic_ordering,wahl2007}. The impurities are coupled
through exchange interactions of different physical origins, either
direct exchange for nearest-neighbor adsorption sites or indirect
substrate-mediated coupling that asymptotically decay as a power-law 
with increasing separation between the impurities
(Ruderman-Kittel-Kasuya-Yosida or RKKY interaction)
\cite{Yoisida.Magnetic_properties,Kittel.Indirect_exchange,Kasuya.Theory_of_metallic}.
The indirect exchange coupling depends on the nature of the substrate
states involved: in addition to bulk states, some commonly used noble
metal substrates, such as Cu(111), also have a band of surface states
crossing the Fermi level \cite{Gartland.Transition,hasegawa1993}. The
existence of surface states in metals was first predicted in 1930s by
I. Tamm and W. Shockley\cite{Tamm,Shockley.surface,DavisonBook}.
Originating from the atomic levels, it was shown how these states
appear when the boundary of the crystal is formed. The
momentum-resolved electronic structure on the surface of
copper\cite{Heimann.Obeservation} and other noble
metals\cite{Reinert.Direct_measurements} has been studied using 
angle-resolved photoemisson spectroscopy
 \cite{Winkelmann.Direct_k_space}. More recently,
the dispersion of surface states has also been studied in Cu and Ag
using an STM\cite{Burgi.Noble_metal_surface}. First principles and
analytic calculations have predicted long-ranged oscillatory
adsorbate-adsorbate interactions mediated by these surface
states\cite{Simon.Exchange_interaction,Patrone.Anisotropic_RKKY,Hyldgaard.Long_ranged}.
Their findings are consistent with a 2D RKKY interaction that decays
asymptotically as $\sim1/R^2$. 

In the presence of both bulk and surface states, as is commonly the
case, the distance dependence of the exchange coupling at intermediate
distances, which are also physically the most relevant, is
non-trivial. Moreover, at very short inter-impurity separation the
atomistic details become important and lead to significant directional
anisotropy. In addition, for strong impurity-substrate couplings the
effective impurity-impurity exchange coupling can no longer be reliably
determined through low-order perturbation theory estimates as in the
simple RKKY picture. This approach become particularly troublesome in
situations where non-perturbative effects, such as the Kondo screening
\cite{ternes2009,Requist:2016em,brune2009}, are also significant.  For
instance, in a recent study of Fe atoms on Cu(111) surface, their
mutual interactions were measured with spin-resolved STM
\cite{Wiesendanger.Atom_by_atom} and the data was interpreted in terms
of an Ising model which does not include any many-body effects. In
such cases, reliable non-perturbative techniques are required for an
unbiased analysis. Such methods, based on the density matrix
renormalization group (DMRG) group, have recently been developed
\cite{Feiguin.Lanczos,Allerdt.Kondo}. They allow one to study the full many-body problem with 
a realistic 
description of the band structure obtained from atomistic first principles calculations.
The resulting tight-binding-model
description of the substrate can be {\it exactly} mapped onto a 1D representation
suitable for the DMRG calculations. This approach is numerically exact and correctly describes
correlation effects and, in particular, the Kondo singlet formation.

Many of the early STM experiments on single magnetic adsorbates
exhibiting a Kondo resonance were performed on the (111) surfaces of
noble metals such as Cu and Ag that host surface-state bands
\cite{li1998,manoharan2000,limot2004}. Since both surface-state and
bulk conduction band electrons hybridize with the adsorbate, the
question of which one plays the dominant role has been widely debated
\cite{knorr2002,barral2004,lin2005,lin2006}. A number of experiments suggested a
more important role of the bulk states in the formation of Kondo state
even on (111) surfaces \cite{schneider2002a,knorr2002,barral2004,limot2004}: for
example, the resonance width is not affected when the adatom is moved
close to the step edges where the surface-state electron local density
of states is modulated by a standing wave pattern \cite{limot2004}. On
the other hand, the quantum mirage experiments \cite{manoharan2000}
clearly demonstrate that the Kondo resonance is projected from one
focus to the other in an elliptical quantum corral, which necessarily
involves the surface-state electrons. Recently, Kondo physics has also
been explored on the Si(111)-$\sqrt{3}\times\sqrt{3}$Ag substrate,
which is a semiconductor with a metallic surface: a long decay length
of the Kondo resonance has been observed \cite{Li:2009fj}.

The usual treatment to derive an effective exchange interaction between the localized moments involves second-order perturbation theory. The result can be summarized as:
\[
J_{RKKY}({\bf R})=J_K^2\chi({\bf R}),
\]
where $\chi({\bf R})$ is just the Fourier transform of the non-interacting static susceptibility, or Lindhard function. 
In this paper, we numerically study the adatom-adatom interactions on a surface
with surface states non-perturbatively and with full real-space resolution. We consider
quantum spins $S=1/2$ and we show important departures from the
conventional perturbative RKKY interpretation. In particular,
long-ranged interactions are absent because of the formation of
separate Kondo-singlet states\cite{Allerdt.Kondo}.

The paper is structured as follows: In Section II we introduce the model and methods,
emphasizing on a new computational development that is introduced in this context for the first time.
We describe our results in Section III, and we finally close with a discussion.

\section{Model and Method}

\subsection{ Band Structure }
\begin{figure}
  \centering
  \includegraphics[scale=0.28]{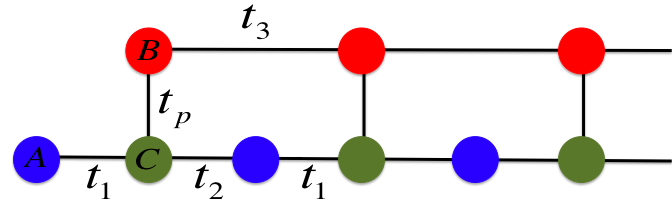}
  \caption{Diagram of the chain with sites A, B, and C and 
  hoppings $t_1, t_2, t_3$ and $t_p$.}
  \label{shockley_chain}
\end{figure}

Unlike the boundary states in topological insulators, the Shockley surface states are metallic and coexist near the Fermi level with other bulk bands that are also partially filled.
We begin our discussion with a three band tight binding model on a
semi-infinite chain. We will then be able to extend the idea to an FCC
lattice, where each site on the chain will become a triangular plane arranged to form an FCC lattice. The bands are described by hopping parameters $t_1, t_2,
t_3$ and $t_p$ as shown in Fig.~\ref{shockley_chain}. This model will
host a surface, or edge, state on the topmost $A$ sites. We follow the
method described by Pershoguba and Yakovenko in
Ref.~\onlinecite{Yakovenko.shockley} to find the energy of this state
by extending the treatment to a three-band problem. 
The Hamiltonian for a translational invariant chain in momentum space can be represented by the matrix:
\begin{equation}
H =
 \begin{pmatrix}
  \epsilon_A & 0 & t_1+t_2e^{ik}  \\
  0 & \epsilon_B+t_3\cos{(k)} & t_p \\
  t_1+t_2e^{-ik}& t_p & \epsilon_C  \\
 \end{pmatrix}
 \end{equation}
 
One key feature of this Hamiltonian that allows us to find an explicit expression for the surface state, is that the $A$ sites are not directly coupled to the $B$ sites. The Schrodinger equation can be expressed as:
\begin{equation}
V\Psi(z) + (U-E)\Psi(z+1) + V^{\dagger}\Psi(z+2)=0
\end{equation}
where we introduce the spinor
\begin{equation}
\Psi(z)=
\begin{pmatrix}
\psi_A(z) \\
\psi_B(z) \\
\psi_C(z)
\end{pmatrix}
\end{equation}
Here $z \ge 1$ labels the unit cell along the chain direction and the wave functions $\psi$ sit on sites $A$, $B$, or $C$, which represent three orbitals of a copper atom that can originate from hybridizations of $s$, $p$, and $d$ atomic orbitals \cite{Harrison2003}.
In this expression we use the matrices $U$ and $V$ defined as:
\begin{equation}
U =
 \begin{pmatrix}
  \epsilon_A &  0 & t_1  \\
  0 & \epsilon_B & t_p \\
  t_1 & t_p & \epsilon_C  \\
 \end{pmatrix}
\qquad V =
 \begin{pmatrix}
  0 &  0 & t_2  \\
  0 & t_3 & 0 \\
  0 & 0 & 0  \\
 \end{pmatrix}~.
\end{equation}
Following Ref.~\onlinecite{Yakovenko.shockley}, we introduce the generating function
\begin{equation}
G(q) = \sum_{z=1}^{\infty} q^{z-1}\Psi(z)~,
\end{equation}
that can be re-written as
\begin{equation}
G(q) = [q^2V+q(U-E)+V^{\dagger}]^{-1}\Psi(1)~,
\end{equation}
where we have used the boundary condition
\begin{equation}
(U-E)\Psi(1) + V^{\dagger}\Psi(2)=0
\end{equation}

Yakovenko {\it et al.}\cite{Yakovenko.shockley} showed that for a
given energy $E$, a surface state exists if all poles of $G(q)$ have
magnitude greater than one. For our 1D chain, this state exists at $E=\epsilon_A$, just as the two-orbital case, under the condition $|t_2| >|t_1|$. Also note that the surface state {\it only} exists on the $A$ sites.

This same procedure can be generalized to a three-dimensional structure (an FCC crystal in our case). The hoppings now acquire an in plane $(p_x, p_y)$ dependence.  The in-plane hoppings are denoted as $ t_A, t_B$ and $t_C$. The Hamiltonian of the bulk is now
\begin{equation}
H =
 \begin{pmatrix}
  \epsilon_A + h_A& 0 & t(\vec{p},k)  \\
  0 & \epsilon_B+h_B+t_3(\vec{p},k) & t_p \\
  t^*(\vec{p},k)& t_p & \epsilon_C + h_C \\
 \end{pmatrix} ~.
 \end{equation}
where,
\begin{gather}
h_\lambda = t_\lambda\left( 2\mathrm{cos}(p_x)+4\mathrm{cos}(p_x/2)\mathrm{cos}(\sqrt{3}p_y/2) \right)  \nonumber \\
t(\vec{p},k)= t_1 + t_2e^{-i\sqrt{2/3}k} \left(e^{-i\sqrt{3}p_y/3}+2e^{i\sqrt{3}p_y/6}\mathrm{cos}(p_x/2)\right)  \nonumber \\
t_3(\vec{p},k) = t_3 \left( 2\mathrm{cos}(\sqrt{2/3}k + \sqrt{3}p_y/3 )\right.  \nonumber \\ 
+  \left. ~4\mathrm{cos}(p_x/2)\mathrm{cos}(\sqrt{2/3}k - \sqrt{3}p_y/6 ) \right) ~,
\end{gather}
where the subindex $\lambda$ can represent either one of the three orbitals $A$,$B$,$C$.
Following the same reasoning as before for a chain, it can be found that the energy of the surface state is given by
\begin{equation}E = \epsilon_A + t_A\left(2\mathrm{cos}(p_x)+4\mathrm{cos}(p_x/2)\mathrm{cos}(\sqrt{3}p_y/2)\right)~,
\end{equation}
which is the same as the dispersion of the triangular lattice.

\begin{figure}
  \centering
  \includegraphics[scale=0.36]{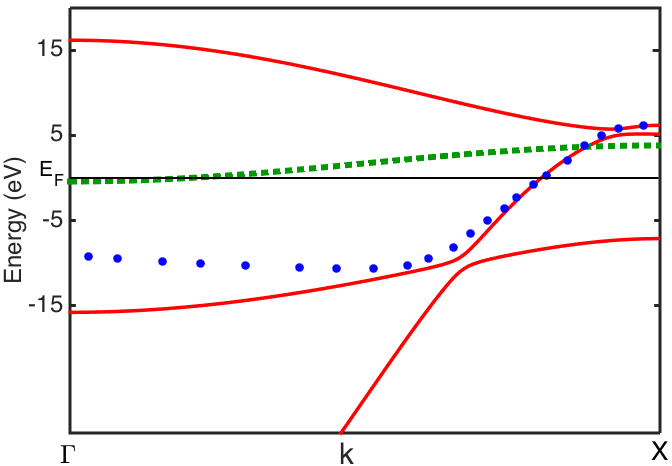}
  \caption{Band structure of our system as obtained with the model described in the text. Red
  lines show bulk bands while the green line is the surface state.
  Blue points are theoretical values taken from Ref.~\onlinecite{Burdick.Copper} }
  \label{bands_shockley}
\end{figure}

\begin{figure}
  \centering
  \includegraphics[scale=0.35]{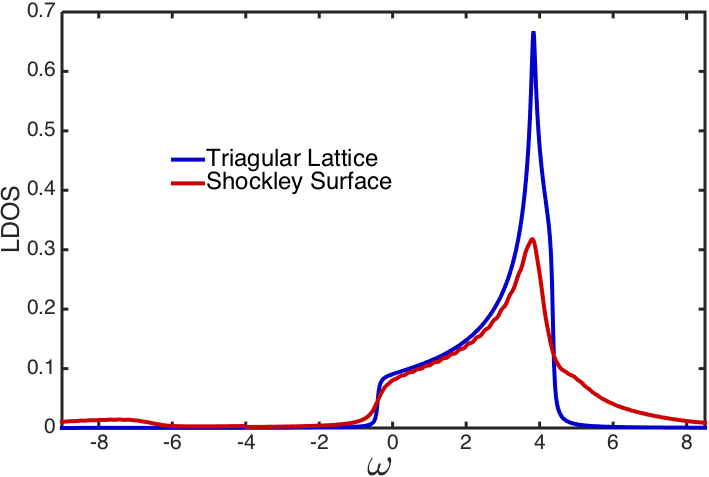}
  \caption{LDOS for a site on the Shockley surface as well as a 2D triangular lattice with the same bandwidth, where zero energy corresponds to the Fermi level. The 2D results are in the thermodynamic limit, while the Shockley data corresponds to 500 poles. Bulk states have very small weight below the Fermi level. }
  \label{shockley_dos}
\end{figure}

The parameters in the model are adjusted to give a good description matching both first principles calculations and experiments. First, we fix $t_A$ and $\epsilon_A$ to fit the surface state energy at the $L$-point and its Fermi momentum. This state has a binding energy of about $-0.4eV$ and Fermi vector $k_f \approx 0.2 \AA^{-1}$ \cite{Winkelmann.Direct_k_space}. The remaining parameters (shown in Table \ref{table}) are used to fit the bands near the Fermi level along the $\Gamma-X$ line (Fig.~\ref{bands_shockley}). 
The theoretical data shown in Fig.~\ref{bands_shockley} is taken from
Ref~.\onlinecite{Burdick.Copper}.  Using this model, we obtain the
correct surface band energy near the $\Gamma$ and $L$ points, we match
the Fermi momentum of the surface band, fit the bulk bands along
$\Gamma-X$ near the Fermi energy, all by preserving the symmetry of an
FCC lattice. The resulting bands are shown in
Fig.~\ref{bands_shockley} as well as the local density of states in
Fig.~\ref{shockley_dos}. Notice the local density of states (LDOS) of the surface has a form
very similar to that of a 2D triangular lattice with some
contributions from the bulk bands, as expected.

\begin{table}
\begin{centering}
\begin{tabular}{| c | c | c | c | c | c | c | c | c | c |}
	\hline
	$t_1$ & $t_2$ & $t_3$ & $t_p$ & $t_A$ & $t_B$ & $t_C$ & $\epsilon_A$ & $\epsilon_B$ & $\epsilon_C$ \\
	\hline
	 1.0 & 5.0 & -4.0 & -1.0 & -0.53 & -4.0 & 0.8 & 2.77 & -10.0 & -4.2 \\
	\hline
\end{tabular}
\caption{Model parameters (in $eV$) used in calculations throughout the paper. Note that small changes in the parameters do not induce qualitative changes in the results.}
\label{table}
\end{centering}
\end{table}

\subsection{Numerical approach}

Having obtained an accurate representation of the bulk and surface
bands, we now introduce the magnetic atoms as two $S=1/2$ Kondo
impurities at positions $r_1$ and $r_2$, connected to the surface
through the many-body exchange interaction:
\begin{equation}
V = J_K \left( \vec{S}_1 \cdot \vec{s}_{\rm r_1} + \vec{S}_2 \cdot \vec{s}_{\rm r_2} \right),
\label{hamiltonian1}
\end{equation}
where $J_K$ is the Kondo coupling constant.

In order to make the problem numerically tractable, we employ the so-called block Lanczos method recently introduced in this context by two of the authors \cite{Allerdt.Kondo,Yunoki.Block}. This approach is inspired by Wilson's original formulation of the numerical renormalization group \cite{Wilson.nrg}, but accounting for the lattice structure. It enables one to study quantum impurity problems in real space and in arbitrary dimensions with the density matrix renormalization group method (DMRG)\cite{White1992,White1993}. This is done through a unitary transformation to a basis where the non-interacting band Hamiltonian has block diagonal form.
As described in detail in Refs.~\onlinecite{Allerdt.Kondo,Yunoki.Block}, this is equivalent to a  block Lanczos iteration, where the recursion is started from seed states corresponding to electrons sitting at the positions of the impurities.
The resulting matrix that can be re-interpreted as a single-particle Hamiltonian on a ladder geometry.

\begin{figure}
  \centering
  \includegraphics[scale=0.3]{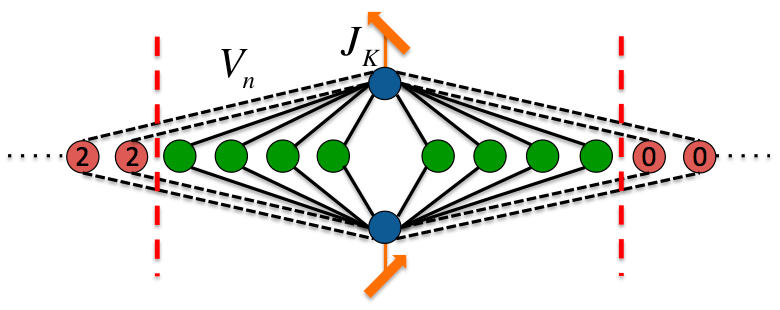}
  \caption{A representation of the star geometry. The blue circles represent sites in real space or the "seed" states in the Lanczos transformation. The new bath sites are shown in green and red. The green represent the relevant sites while the red are discarded indicated by the dashed red line where the system is effectively truncated. The red sites are either double occupied ("2") or empty ("0"). The orange arrows correspond to Kondo spin-1/2 impurities coupled via $J_K$.}
  \label{star}
\end{figure}

The Lanczos transformation is carried out by taking $A$ sites on the
surface at the impurity positions to be the seed states. 
Since we are considering impurities on the surface of a 3D system, a
ladder of length $L$ will contain contributions from both surface and bulk states, with the large majority of these states having larger weight in the bulk. 
In order to reach the desired extremely large systems, we introduce an improvement to the afore mentioned method. It was shown that transforming from a ladder to a star geometry results in lower ground state entanglement than the ladder geometry\cite{Wolf.Solving_nonequilibrium}, making it ideal for DMRG. A pictorial representation of the new geometry for the two impurity problem is shown in Fig. ~\ref{star}. This mapping corresponds to a second unitary transformation on top of the ladder geometry. The seed states remain unchanged and are now coupled to the bath states with on-site energy $\epsilon_n$ via new hoppings $V_n$. The new system is very advantageous since the surface states are very weakly coupled to the bulk states at high and low energies. We find that high energy states are either doubly occupied or empty and can be discarded without losing any physics. This allows us to carry out simulations in extremely large systems of the order of $1600^3$ sites keeping of the order of 200 orbitals! We find that the results are absolutely free of finite size effects and therefore they represent the thermodynamic limit for all practical purposes.
In all our simulations, we use $1000$ DMRG states, which yield results with machine precision accuracy in both energy and correlations.

\section{Results}
We first recall the expression for the Lindhard function:
\begin{equation}
\chi(\mathbf{r}_1,  \mathbf{r}_2)=2\mathrm{Re}\sum\frac{\braket{\mathbf{r}_1 | n}\braket{n|\mathbf{r}_2}\braket{\mathbf{r}_2|m}\braket{m|\mathbf{r}_1}}{E_n-E_m},
\end{equation}
where the sum is over the eigenstates $n,m$ with energies $E_n>E_F>E_m$.
The $\ket{\mathbf{r}_{1,2}}$ are the single-particle states at positions $\mathbf{r}_{1,2}$.
This function can be computed numerically for our systems, and compared to the non-perturbative results obtained by solving the many-body problem.

\begin{figure}
  \centering
  \includegraphics[scale=0.44]{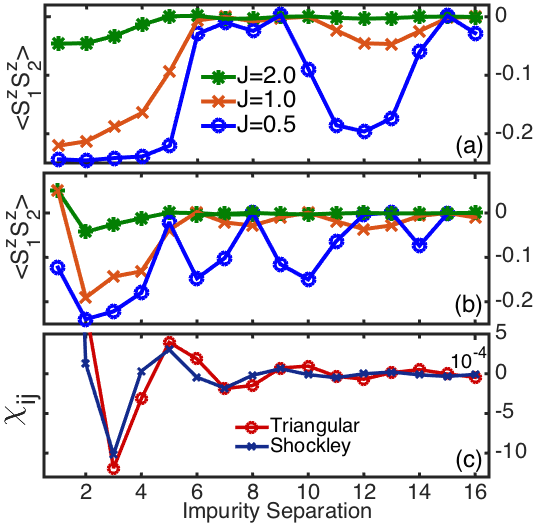}
  \caption{Spin-spin correlations as a function of the impurity
  separation for (a) triangular lattice and (b) 2D Shockley surface of a (111) metal. (c) Lindhard function for the corresponding lattices. All plots are along the nearest neighbor direction.}
  \label{shockley_corr}
\end{figure}

Results for spin-spin correlations as well as the perturbative result (Lindhard function) are shown in Fig.~\ref{shockley_corr}. We only display the $z$ component of the spin since the problem is $SU(2)$ symmetric. The particle number was adjusted to match the Fermi level in copper. For the triangular lattice, the filling is such that the Fermi level is at the same point as the surface band, i.e., we are in the low density regime. An interesting feature is that ferromagnetism is found on the Shockley surface only at $R=1$, consistent with the Lindhard function as well as the result in Ref.~\onlinecite{Stepanyuk.Magnetic_nano}. Beyond $R=1$ however, the Kondo effect will dominate over ferromagnetism, where the perturbative results predict oscillations. 

In contrast, on the triangular lattice we find that all correlations are anti-ferromagnetic regardless of the value of $J_K$.  At weaker coupling, the impurities transition into a free moment regime attributed to the very low carrier density. When this occurs, our correlations acquire the value $\langle S_1^zS_2^z\rangle=-1/4$, the magnetic moments are completely decoupled from the conduction electrons and from each other, and the ground state is 4-fold degenerate with spins pointing in either direction. Since we are enforcing spin conservation and $S^z_\mathrm{Tot}=0$, the impurities are always anti-parallel and correlations can only assume the value $-1/4$. As the coupling is increased to $2eV$, the spins form their own Kondo clouds after a separation of just a few lattice spacings, signaled by zero correlations. In this regime, both staggered and uniform magnetic susceptibilities are the same, and equal to the single impurity case. 
Calculations were also conducted with increased filling on the
triangular lattice, and our results are qualitatively the same, {\it
i.e.}, there is a competition between anti-ferromagnetism and Kondo,
with ferromagnetism completely absent. This is contrasting to the
square lattice\cite{Allerdt.Kondo} and graphene\cite{Allerdt.Graphene}
where some ferromagnetism is found at half filling, and is here due to the non-bipartite nature of the triangular lattice.

There is a striking difference in the correlations on the Shockley surface as a function of $J_K$. 
At lower values of the coupling $J_K$ correlations for the Shockley
surface and the triangular lattice differ noticeably, but as the interaction is increased, they start resembling each other with the first one having a slight increase in the screening due to the contributions from the bulk states. 
Not only is there an increase in Kondo screening, but there is a significant change in the wavelength of oscillation. This change in wavelength is also seen in the Lindhard function, but it is not as dramatic as the many-body case. We attribute these effects to the influence of the bulk. The periods of the oscillations are similar to the the periods of the Lindhard function, except at short distances where the impurities are tending toward free moments. 
Remarkably, at very short distances the correlations actually acquire the opposite sign! 
For larger $J_K$, and $R=1$, the Lindhard function correctly predicts the ferromagnetic interaction for the Shockley case. This behavior is not seen in systems where there is only one band present at the Fermi level. 

In addition to placing both impurities on the surface, calculations were done with one impurity in the bulk. It is found that correlations vanish after just one lattice space regardless of $J_K$, indicating that the surface states do not leak into the bulk: The impurity at the surface will couple mainly to the surface states (as indicated by the LDOS in Fig.~\ref{shockley_dos}), and the impurity in the bulk will couple to predominately bulk states since the surface states decay exponentially into the bulk. As a consequence, the impurities will remain uncorrelated.

\begin{figure}
  \centering
  \includegraphics[scale=0.40]{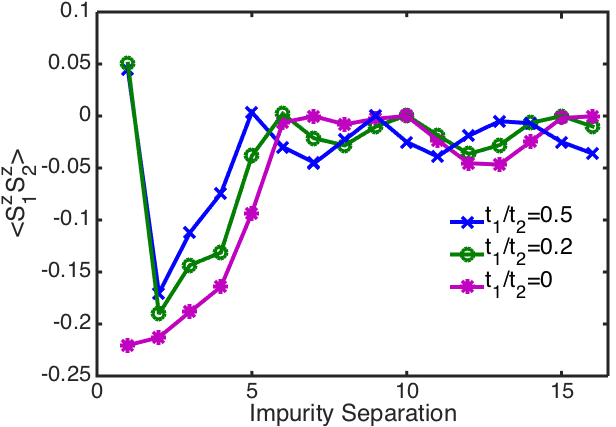}
  \caption{Spin-Spin correlations for different values of ${t_1}/{t_2}$ for two spin $S=1/2$ impurities on the Shockley surface with $J_K$=1.0.}
  \label{t2t1}
\end{figure}

In Fig. \ref{t2t1} we show results obtained from different values of $\alpha=t_1/t_2$. The case $\alpha=0$ corresponds to a completely decoupled surface, the triangular lattice. As $\alpha$ is increased, the bulk contributes more and more to the physics. This can be seen as a change in the period of the oscillations and the sudden and dramatic change of sign of the correlations at distance $R=1$.

\section{Conclusions}
We have investigated the role of surface and bulk states on the effective indirect exchange between two quantum impurities. The metallic surface was modeled by means of a tight-binding Hamiltonian that reproduces the surface and bulk bands near the Fermi surface, as obtained from first principles calculations. The quantum many-body problem was then solved numerically by means of the DMRG method after mapping the non-interacting degrees of freedom to a star geometry. This approach allows us to study the problem in the thermodynamic limit, with machine precision accuracy. 
Most remarkably, ferromagnetism is completely absent, with the exception of impurities at distance $R=1$, departing from the behavior observed in an isolated 2D triangular lattice. This effect clearly illustrates the subtle artifacts of perturbative approaches that are usually employed to guide experiments. We point out that at short distances it is very likely that direct exchange due to overlaping wave functions dominates the physics. 

We found that the correlations extend to just one atomic layer into the bulk, and the main contributions originate from surface states. 
However, our results indicate that the bulk states introduce a change in the period of the oscillations of the RKKY interaction, and a non-trivial competition with Kondo physics. 
The RKKY interaction in these systems cannot be completely described in terms of an isolated 2D surface as also observed experimentally for the case of a single impurity in the Kondo regime \cite{schneider2002a,knorr2002,barral2004,limot2004}.

\begin{acknowledgments}
We are grateful to C.~F. Hirjibehedin for useful discussions. AA and AEF acknowledge the U.S. Department of Energy, Office of Basic Energy Sciences, for support under grant DE-SC0014407. 
R\v{Z} acknowledges the support of the Slovenian Research Agency (ARRS) under P1-0044 and J1-7259,
\end{acknowledgments}


\end{document}